\documentclass[fleqn,10pt]{olplainarticle}

\usepackage{verbatim}
\usepackage{float}
\usepackage[%
    colorlinks,
    allcolors=black,
    breaklinks=true,
]{hyperref}
\usepackage{ragged2e}
\usepackage[super,numbers]{natbib}
\usepackage{wrapfig}
\title{Demographic Factors Associated with Triage Acuity, Admission and Length of Stay During Adult Emergency Department Visits}

\author[1,2]{Helena Coggan}
\author[3]{Pradip Chaudhari}
\author[4]{Yuval Barak-Corren}
\author[2,5]{Andrew M. Fine}
\author[1,2]{Ben Y. Reis}
\author[6]{Jaya Aysola}
\author[1,2]{William G. La Cava}

\affil[1]{Computational Health Informatics Program, Boston Children's Hospital, Boston, MA, USA}
\affil[2]{Harvard Medical School, Boston, MA, USA}
\affil[3]{Division of Emergency and Transport Medicine, Children's Hospital Los Angeles and Department of Pediatrics, Keck School of Medicine of the University of Southern California, Los Angeles, CA, USA}
\affil[4]{Department of Pediatric Cardiology, Schneider Children's Medical Center, Affiliated to Tel Aviv University Faculty of Medical and Health Sciences, Petach Tikvah, Israel}
\affil[5]{Division of Emergency Medicine, Boston Children's Hospital, Boston, MA, USA}
\affil[6]{Leonard Davis Institute of Health Economics, University of Pennsylvania; Department of Medicine, Perelman School of Medicine, University of Pennsylvania; and Penn Medicine Center for Health Equity Advancement, Philadelphia, Pennsylvania}


\begin{abstract}
\textit{\textbf{Objective:} To describe the association of demographic factors with triage acuity, hospital admission rates, and length of stay (LOS) for adult patients in the emergency department (ED).\newline
\textbf{Methods:} 
We performed a retrospective cross-sectional analysis using publicly available electronic health records describing visits to the ED of a single US medical center during 2011-2019. The primary exposures of interest were self-reported gender, race/ethnicity, and age. The outcomes studied were triage acuity, admission to hospital, and LOS in the ED. Odds ratios were calculated using propensity-score matching. Analyses were adjusted for confounding variables, including vital signs and diagnoses.
 \newline
\textbf{Key Results:} Black patients were more likely than White patients to experience long stays before admission but not before discharge. Men were more likely than women to be triaged as urgent or admitted, and had shorter stays. Patients over 30 were likelier to be triaged as urgent or admitted, but had longer stays.}
\end{abstract}

\begin{document}

\flushbottom
\maketitle
\thispagestyle{empty}
\justifying
\section{Introduction}\label{sec:introduction}

Persistent disparities have been observed in the quality of healthcare received by patients from different demographic groups \cite{Nelson2002}, including in the emergency department (ED) setting \cite{Gabayan2013, Longcoy2022, Martin2024}. These inequities are particularly stark in patients from marginalised racial groups: for example, Black patients are more likely than White patients to experience extended stays before admission to hospital, in both adult \cite{Aysola2022, Lu2021, Pines2009} and pediatric contexts \cite{James2005, Park2009}. Research has also consistently shown that Black patients are assigned lower triage scores than White patients, receive less thorough medical investigations, and wait longer to see a physician \cite{Sonnenfeld2012, Schrader2013, Joseph2023}. Black and Hispanic race, Medicaid insurance, and a preference for Spanish as a primary language have also been established as risk factors for deprioritisation in the ED \cite{Sangal2023}.

Although the factors driving these care disparities are not fully understood, they may be partly driven by implicit provider bias, as shown by recent mixed-methods work finding that physicians are less likely to communicate well when their race differs from that of the patient \cite{Aysola2022}. 
Women are also more likely than men to report difficulties in obtaining timely care in the ED \cite{Chen2022}. 

Outcome inequalities on the basis of protected characteristics such as gender, race and age are especially dangerous in the context of emergency medicine, where they can have immediate and life-changing consequences. Efforts to address these disparities first require them to be carefully quantified. We investigated differences in ED outcomes between demographic groups using a large, publicly available single-site dataset. 

In doing so, we sought to address a number of key gaps in the literature. Previous studies of racial and ethnic disparities in the US healthcare system have generally focused on comparing White, Black and Hispanic patients, lacking the sample size to investigate disparities in other ethnic groups. Our dataset contained more than 17,000 visits by Asian patients, allowing us to investigate whether differences in outcome observed amongst other patients of color extended to this group. We also sought to examine whether the effects of gender, race and age on rates of admission and LOS persisted when we accounted for changes in a patient's condition after triage, a factor not included (to our knowledge) in previous studies. When measuring differences in admission rates and LOS, we also accounted both for a patient's chief complaint (i.e. the nurse's initial assessment of the reason for visit) and for the actual diagnosis eventually attached to their ED stay, to account for differences in a patient's ability to communicate their condition; previous work has generally included one or the other, which may not give an accurate picture of what is known at various points about a patient's condition. We investigated whether patients from marginalised groups waited longer to be discharged, as opposed to merely waiting longer to be admitted, as longer wait times to discharge may be associated with a loss of trust in healthcare systems and dissuade patients from seeking prompt care when necessary. To our knowledge this is the first study of its kind to consider discharge times, and the first to consider outcomes of acuity, admission and LOS in parallel. Finally, as previous studies have investigated patterns of disparity in non-public datasets \cite{Abdulai2021, Aysola2022, Gabayan2013, Sangal2023, Joseph2023, Sonnenfeld2012, Schrader2013} or in a single dataset without information on vital signs \cite{Lu2021,Abdulai2021,Park2009}, we examined a newly-available, public single-site dataset, MIMIC-IV-ED \cite{johnsonalistairMIMICIVED2021}.

Healthcare disparities observed in such datasets have the potential to be reproduced by machine learning and artificial intelligence (AI) models trained on these data for clinical decision support~\cite{lacavaFairAdmissionRisk2023a}.  
Several data-driven AI tools for the ED already exist, for example to predict admissions~\cite{Barak-Corren2017,Barak-Corren2021} and clinical orders~\cite{hunter-zinckPredictingEmergencyDepartment2019}, which in turn may improve patient flow and case management.
Given the high potential for current and future models to leverage readily available ED data, we sought to address our research questions using MIMIC-IV-ED~\cite{johnsonalistairMIMICIVED2021,Goldberger2000}.  
This is a large, publicly available database of visits by adult patients to the ED of Beth Israel Deaconess Medical Center, a tertiary academic medical center in Boston, Massachusetts. 
It is vital that researchers designing these tools are aware of possible sources of bias within the data itself, to ensure that they do not reinforce existing inequalities. 
This is especially true in the light of the final rule on Section 1557 of the Affordable Care Act, that holds health providers responsible for discrimination arising from their use of AI-based decision support interventions~\cite{departmentofhealthandhumanservicesNondiscriminationHealthPrograms2024}.

\section{Methods and Materials}\label{sec:methods}

\subsection{Data Sources and Study Design}

We conducted a single-center retrospective cross-sectional study of ED visits by adults to BIDMC. Visits were excluded if they did not have a valid recorded `in-time' (time of arrival to the ED) or `out-time' (time at which they left the ED); to be valid, an in-time had to precede the recorded out-time. We obtained vital signs taken at triage, and excluded visits for which such data was not available, or where vitals fell outside a reasonable range (e.g. a temperature of 200 degrees Farenheit). We obtained vital signs recorded during a patient's stay, but included stays without such information, as many patients may be admitted too quickly to allow repeated vital assessments. 

Age was obtained to the nearest year by linkage to MIMIC-IV, assuming a birthday of January 1st for each patient. We converted ICD codes relating to Charlson comorbidity scores \cite{Quan2011} (a measure of severity) using the R package `comorbidity' \cite{Gasparini2018}. Visits were excluded if no ICD code was available; linked ages were observed for all patients. In total, 35,252 visits, or 8.3\% of those in the database, were excluded due to missing or incomplete information.

Patient-reported pain, generally reported as a number between 0 and 10, was converted to a score within this range (i.e. interpreting 90 as out of 100 and thus converting it to 9, or `flipping a coin' to convert 6.5 to either 6 or 7). Pain values corresponding to `unable to answer' were retained as a separate category of response. Chief complaints were converted to 54 manually curated binary indicators covering the most frequent reasons given for a visit to the ED, such that 88\% of visits were described by at least one variable.

Visits resulting in an admission to hospital were assigned a specific linkage ID which allowed us to obtain data on the patient's insurance status, marital status, and primary language from the MIMIC-IV dataset. These data were not available for visits which did not result in an admission. As a result, we were only able to control for these variables when examining demographic disparities in LOS amongst those admitted to the hospital (see below).

\subsection{Exposures of Interest}

Our exposures of interest included gender (female or male), race (Asian, American Indian/Alaska Native, Black/African-American, Native Hawaiian/Pacific Islander, Hispanic/Latino, White, other, and unknown), and age (18-29, 30-39, 40-49, 50-59, 60-69, 70-79, 80-89, and 90+). Exposures were recorded at the time of ED visit and we did not exclude patients whose reported gender or race varied between visits. 

For patients admitted to hospital, we also considered insurance status (Medicare, Medicaid, private, or other/NA) and primary language (English or other) as exposures.

\subsection{Outcome Measures}

We considered two primary outcomes: 1) whether or not a patient was triaged as urgent (ESI score of 1 or 2), and 2) whether a patient was admitted to hospital. We also considered whether a patient experienced a `long stay', defined as an LOS greater than the relevant median, where medians were calculated separately for patients admitted to hospital and patients discharged to their homes (resulting in `high LOS' thresholds of 390 minutes and 300 minutes respectively). Patients with other outcomes (e.g. those who left without being seen) were excluded from LOS analysis. As a sensitivity analysis, we defined a `very long stay' as a length of stay in the top quartile (i.e. longer than 540 minutes for admitted patients, and longer than 465 minutes for discharged patients). For LOS analyses, we excluded visits with a stay longer than 24 hours, to avoid confounding by possible recording errors.

\subsection{Adjustment Variables}

When assessing the influence of demographic factors on triage urgency, we adjusted for chief complaint, patient-reported pain (0, 1-3, 4-7, 8-10, unable to answer, or non-numeric response), mode of arrival at hospital (ambulance, helicopter, walk-in, unknown or other), and time of arrival at the ED (00:00-05:59, 06:00-11:59, 12:00-17:59, or 18:00-23:59). We also adjusted for associated year-group (2008-2010, 2011-2013, 2014-2016, or 2017-2019), to account for temporal changes in triage procedures or hospital capacity. We included two additional variables indicating whether or not a patient had visited the hospital in the last 30 days, with or without a resulting admission. We also controlled for vital signs taken at triage: temperature (above or below 100.4F, indicating the presence or absence of a fever); low, normal or high respiration rate (low: \(<12 \) breaths per minute, high: \(> 20 \) bpm), low, normal, high or very high heart rate (low: \(<60\) beats per minute, high: \(100-150 \) bpm, very high: \(> 150\) bpm), and blood pressure (low, normal, elevated, stage 1 high, stage 2 high, or unstably high). Each of these categories was encoded as a binary dummy variable with respect to a specified reference category (the `normal' reading): for instance, respiration rate was represented by two binary variables, `resp\_low' and `resp\_high', with zeros in both variables indicating that the patient's respiration rate was normal at triage.

When the outcome under consideration was not triage urgency (i.e. admission to hospital or LOS), we further adjusted for variables not known to the triage nurse. These include the actual triage acuity assigned (ESI score 1, 2, 3, or 4/5); Charlson comorbidity score, a measure of the severity of the actual diagnosis resulting from the ED stay (0, 1, or 2+); and vitals taken during the ED visit. These vitals were filtered and categorised identically to the vitals taken at triage, and encoded as a series of binary indicators. Each dummy variable indicated whether or not an abnormal reading was \textit{ever} recorded during a patient's stay (i.e. a value of `resp\_high'=1 would indicate that a patient's respiration rate was abnormally high during at least one reading). We included additional indicators, one per category of vital sign (heart rate, blood pressure, temperature, and respiration rate), denoting whether there were any readings during a patient's stay in the course of which a particular vital was \textit{not} taken, on the grounds it may be informative when a nurse does not consider a particular vital worth measuring.

Records of visits resulting in a hospital admission also contained linked information on insurance status (Medicare, Medicaid, private, or `other', a category including those not charged and those for which insurance information was not available), marital status (single, married, widowed, divorced, and N/A), and primary language (English or other). We adjusted for these variables when the outcome was a long stay before admission, but could not adjust for them when the outcome was triage acuity, admission, or a long stay before discharge.

\begin{table}[h]
\centering
\footnotesize
\caption{\label{tab:characteristics} Demographic characteristics of the study population in the MIMIC-IV-ED dataset. Numbers correspond to visits. P-values are taken from unadjusted chi-squared tests indicating whether a particular demographic variable is associated with admission, relative to its absence (e.g. comparing being Black to not being Black). American Indian includes Alaska Native; Native Hawaiian includes Pacific Islander; Black includes African-American. Significance values were reported with Bonferroni correction.}
\begin{tabular}{ll|cccccc}
 && \multicolumn{2}{c}{Admitted}& \multicolumn{2}{c}{Not admitted}& P&Significance\\\hline
 Gender&Female& 71430& (51.6\%)& 141503&(56.3\%)& \(<0.001\)&***\\
 &Male& 66896& (48.4\%)& 110006&(43.7\%)& \(<0.001\)&***\\
  && & & && &\\
  Race&White& 94154& (68.1\%)& 132126&(52.5\%)& \(<0.001\)&***\\
  &Asian& 5127& (3.7\%)& 12196&(4.8\%)& \(<0.001\)&***\\
  &Black& 23319& (16.9\%)& 63329&(25.2\%)& \(<0.001\)&***\\
  &Hispanic/Latino& 7632& (5.5\%)& 24698& (9.8\%)& \(<0.001\)&***\\
  &American Indian& 375& (0.27\%)& 574&(0.23\%)& 0.01&\\
  &Native Hawaiian& 144& (0.1\%)& 289&(0.11\%)& 0.358&\\
  &Other& 5938& (4.3\%)& 15473& (6.2\%)& \(<0.001\)&***\\
  &Unknown& 1637& (1.2\%)& 2824&(1.1\%)& 0.092&\\
  && & & && &\\
  Age&18-29& 9023& (6.5\%)& 61884&(24.6\%)& \(<0.001\)&***\\
  &30-39& 10579& (7.6\%)& 40393& (16.1\%)& \(<0.001\)&***\\
  &40-49& 13835& (10\%)& 35417&(14.1\%)& \(<0.001\)&***\\
  &50-59& 23794& (17.2\%)& 42816&(17\%)& 0.16&\\
  &60-69& 28255& (20.4\%)& 33060&(13.1\%)& \(<0.001\)&***\\
  &70-79& 24905& (18\%)& 20912&(8.3\%)& \(<0.001\)&***\\
  &80-89& 19874& (14.4\%)& 12622&(5\%)& \(<0.001\)&***\\
  &90+ & 8061& (5.8\%)& 4405&(1.8\%)& \(<0.001\)&***\\
 & & & & & & &\\
 Triage acuity& Urgent (ESI 1/2)& 79537& (57.5\%)& 64814& (25.8\%)& \(<0.001\)&***\\
 & Moderate (ESI 3)& 58209& (42.1\%)& 158561& (63\%)& \(<0.001\)&***\\
 & Low (ESI 4/5)& 539& (0.39\%)& 28019& (11.1\%)& \(<0.001\)&***\\
 & N/A& 41& (0.03\%)& 115& (0.05\%)& 0.02&\\
 \hline
\end{tabular}
\end{table}

\subsection{Data Analysis}

Adjusted odds ratios were calculated using propensity score matching, relative to a specified reference group. References were chosen to be the largest category in each set of exposures (i.e. being female, White, 18-29, speaking English as a primary language, or being a Medicare beneficiary). For each exposure (e.g. Black race), the dataset was limited to visits from patients in the exposed and reference groups (i.e. only visits from Black and White patients). Propensity scores were fitted using a multivariable logistic regression model on all relevant adjustment variables. (These covariates excluded the chief-complaint marker for pregnancy when the exposure was age, and excluded both pregnancy and vaginal bleeding when the exposure was male gender.) Matching was performed using the R package `matchit', with a caliper threshold of 20\% of the standard deviation of the fitted propensity scores. Covariate balance (i.e. an absolute standard mean difference of 0.1 or below for all covariates \cite{Austin2011}
) of was achieved for almost all analyses. Imbalance in some variables was sometimes observed during LOS analysis for visits resulting in an admission. This occurred when the exposure was American Indian race, Native Hawaiian race, or an age of 80 or over, due to sample size constraints. Odds ratios for matched `exposed' and `unexposed' patients were calculated using McNemar's test, a statistical test appropriate to the analysis of paired data. As a sensitivity analysis, we also calculated odds ratios from unmatched data, using propensity scores as an adjustment variable. We found that the results from matching were almost uniformly more conservative (i.e. less likely to be significant), and so to ensure robustness we report only results obtained under this framework. When reporting significance values, we used Bonferroni correction to account for false discovery. Code is available at \url{https://github.com/hcoggan/mimic-iv-ed}
.

\section{Results}
\label{sec:results}

\subsection*{Characteristics of Study Subjects}

Characteristics of the study population are recorded in Table \ref{tab:characteristics}. 

The dataset included 389,835 visits from 190,578 unique patients, of which 138,326 resulted in an admission (with an average admission rate of 35.\%). Of visits which did not result in an admission (\(n=251,509\)), 91\% recorded a disposition of `home' (\(n=229,191\)); we refer to these patients as having been `discharged' (as opposed to merely not admitted). The remainder were recorded as having eloped (\(n=5, 359\)), expired (\(n=61\)), left against medical advice (\(n=1,749\)), left without being seen (\(n=5,826\)), transferred (\(n=6,436\)), or having experienced an unspecified other outcome (\(n=2,887\)). The average length of stay (LOS) was 7 hours and 31 minutes for those admitted, and 6 hours and 53 minutes for those discharged (\(p < 0.001\), two-sided \textit{t}-test). 

In initial unadjusted analyses, we found that male gender, White or American Indian race, an age of 60 or over, and an ESI score of 1 or 2 are factors significantly positively associated with admission. Factors negatively associated with admission include female gender; Asian, Black, Hispanic/Latino, or `other' race; an age under 50; and an ESI score of 3 or lower.

\subsection{Factors Associated With Triage Urgency and Admission}

We calculated adjusted odds ratios for the effects of demographic characteristics on triage acuity (see Table \ref{tab:acuity-admission-ORs}, left column). We found that men were significantly more likely than women to be triaged as urgent (aOR 1.16, 95\% CI 1.14-1.18).

\begin{table}[h]
\centering
\footnotesize
\caption{\label{tab:acuity-admission-ORs} Adjusted odds ratios describing the effect of demographic variables on urgent triage (ESI score 1 or 2) and admission to hospital, relative to a specified reference group. Odds ratios for admission are adjusted for the acuity actually assigned (see text). Significance codes: \(p<0.05 (*); p < 0.01 (**); p < 0.001 (***)\).}
\begin{tabular}{ll|cccc}
 && \multicolumn{2}{c}{Triaged as urgent}& \multicolumn{2}{c}{Admitted to hospital}\\\hline
 Gender&Female& 1& NA (ref)& 1&NA (ref)\\
 &Male& 1.16 (1.14-1.18)& ***& 1.16 (1.14-1.18)&***\\
  && & & &\\
  Race&White& 1& NA (ref)& 1&NA (ref)\\
  &Asian& 0.89 (0.85-0.93)& ***& 0.87 (0.83-0.92)&***\\
  &Other& 0.87 (0.83-0.9)& ***& 0.82 (0.78-0.86)&***\\
  &Unknown& 1.11 (1.01-1.22)& & 1.04 (0.95-1.15)& \\
  &Black& 0.76 (0.75-0.78)& ***& 0.74 (0.73-0.76)&***\\
  &Hispanic/Latino& 0.79 (0.76-0.82)& ***& 0.68 (0.65-0.7)&***\\
  &American Indian& 1.15 (0.94-1.4)& & 1.07 (0.88-1.3)& \\
  &Native Hawaiian& 1.02 (0.76-1.37)& & 0.81 (0.6-1.1)&\\
  && & & &\\
  Age&18-29& 1& NA (ref)& 1&NA (ref)\\
  &30-39& 1.24 (1.2-1.27)& ***& 1.34 (1.3-1.39)& ***\\
  &40-49& 1.37 (1.32-1.41)& ***& 1.76 (1.69-1.83)&***\\
 & 50-59& 1.58 (1.53-1.63)& ***& 2.3 (2.22-2.39)&***\\
 & 60-69& 1.86 (1.8-1.93)& ***& 2.97 (2.85-3.1)&***\\
 & 70-79& 2.08 (1.99-2.17)& ***& 3.56 (3.38-3.75)&***\\
 & 80-89& 2.15 (2.04-2.27)& ***& 4.27 (4-4.57)&***\\
 & 90 +& 2.11 (1.95-2.29)& ***& 4.24 (3.83-4.71)&***\\
 \hline
\end{tabular}
\end{table}

Black, Asian, Hispanic and `other' patients were significantly less likely than White patients to receive a high triage score (Black race aOR 0.76, 95\% CI 0.75-0.78; Asian race aOR 0.89, 95\% CI 0.85-0.93; Hispanic/Latino race aOR 0.79, 95\% CI 0.76-0.82; Other race aOR 0.87, 95\% CI 0.83-0.90). 

When the effects of gender were stratified by race, we found that the effect of male gender was stronger for White and Hispanic men than for Black or Asian men (Asian men vs. Asian women, aOR 1.10, 95\% CI 1.02-1.18; Black men vs. Black women, aOR 1.13, 95\% CI 1.09-1.17; Hispanic men vs. Hispanic women, aOR 1.24, 95\% CI 1.17-1.31; White men vs. White women, aOR 1.17, 95\% CI 1.15-1.20). We found that increasing age up to 70 was strongly associated with triage urgency, and that all patients older than 30 were more likely to be triaged as urgent than those aged 18-29 (see Table \ref{tab:acuity-admission-ORs}, left column).

We next calculated the aORs for admission to hospital, adjusting for triage score, Charlson comorbidity score, and vitals taken during stay. Despite these adjustments, we found significant effects associated with race, gender and age. Men were more likely than women to be admitted to hospital (aOR 1.16, 95\% CI 1.14-1.18), and Black, Asian, Hispanic and `other' patients significantly less likely than White patients to be admitted (Black race aOR 0.74, 95\% CI 0.73-0.76; Asian race aOR 0.87, 95\% CI 0.83-0.92; Hispanic/Latino race aOR 0.68, 95\% CI 0.65-0.70; Other race aOR 0.82, 95\% CI 0.78-0.86). `Unknown' race had no effect on admission likelihood relative to White race. 

When the effects of gender were stratified by race, we again found that the admission effect of being male was stronger for Black, Hispanic and Asian men than White men (Asian men vs. Asian women, aOR 1.24, 95\% CI 1.15-1.33; Black men vs. Black women, aOR 1.18, 95\% CI 1.14-1.22; Hispanic men vs. Hispanic women, aOR 1.22, 95\% CI 1.15-1.29; White men vs. White women, aOR 1.13, 95\% CI 1.10-1.15). We found that the effect of age on admission was stronger than its effect on triage acuity, and that every decade of age up to 80 was associated with an increase in the likelihood of admission, even when the patient's actual condition had been adjusted for as much as possible (see Table \ref{tab:acuity-admission-ORs}, right column).

\begin{table}[h]
    \centering
\footnotesize
    \caption{\label{tab:wait-time-50} Adjusted odds ratios describing the effect of demographic variables on the likelihood of a long stay (greater than the median) before admission (left) or discharge (right), relative to a specified reference group. P-values were adjusted using Bonferroni correction. Significance codes: \(p<0.05 (*); p < 0.01 (**); p < 0.001 (***)\).}
    \begin{tabular}{lc|cccc}
 & & \multicolumn{4}{c}{Long stay before}\\
          &&  \multicolumn{2}{c}{Admission to hospital}&  \multicolumn{2}{c}{Discharge from ED}\\\hline
          Gender&Female&  1&  NA (ref)&  1& NA (ref)\\
          &Male&  0.91 (0.88-0.93)&  ***&  0.9 (0.88-0.91)& ***\\
          &&  &  &  & \\
          Race&White&  1&  NA (ref)&  1& NA (ref)\\
          &Asian&  1.03 (0.94-1.12)&  &  0.94 (0.89-0.99)& \\
          &Other&  1.03 (0.95-1.11)&  &  1.03 (0.98-1.08)& \\
          &Unknown&  0.76 (0.61-0.95)&  &  1.03 (0.85-1.24)& \\
          &Black&  1.14 (1.1-1.18)&  ***&  0.99 (0.97-1.02)& \\
  &Hispanic/Latino& 1.06 (0.98-1.14)& & 1.02 (0.98-1.06)&\\
  &American Indian& 0.88 (0.64-1.2)& & 1.21 (0.94-1.57)&\\
  &Native Hawaiian& 0.5 (0.27-0.9)& & 1.07 (0.73-1.59)&\\
  && & & &\\
  Age&18-29& 1& NA (ref)& 1&NA (ref)\\
  &30-39& 1.13 (1.03-1.23)& & 1.33 (1.27-1.39)&***\\
  &40-49& 1.19 (1.08-1.3)& *& 1.67 (1.58-1.75)&***\\
          &50-59&  1.12 (1.01-1.23)&  &  1.76 (1.68-1.85)& ***\\
 & 60-69& 1.1 (0.98-1.23)& & 1.89 (1.79-2.01)&***\\
 & 70-79& 1.15 (0.96-1.37)& & 2.13 (1.98-2.3)&***\\
 & 80-89& 1.14 (0.89-1.46)& & 2.35 (2.12-2.6)&***\\
 & 90& 1.31 (0.89-1.94)& & 2.29 (1.92-2.73)&***\\
 & & & & &\\
 Primary language& English& 1& NA (ref)& -&-\\
 & Other& 1.08 (1.04-1.15)& & -&-\\
 & & & & &\\
 Insurance& Medicare& 1& NA (ref)& -&-\\
 & Medicaid& 1.04 (0.99-1.08)& & -&-\\
 & Private& 0.95 (0.91-0.99)& & -&-\\
 & Other/NA& 0.87 (0.79-0.95)& & -&-\\
 \hline
    \end{tabular}
\end{table}

\subsection{Factors Associated with Long Stay}

Next, we examined the effect of demographic variables on patient LOS, stratified initially by whether a visit resulted in an admission (Table \ref{tab:wait-time-50}, left column) or a discharge home (Table \ref{tab:wait-time-50}, right column). We initially defined a `long stay' as an LOS greater than the relevant median.

Men were significantly less likely than women to experience a long stay, whether admitted (aOR 0.91, 95\% CI 0.89-93) or discharged (aOR 0.90, 95\% CI 0.88-0.91). When the effect of gender were stratified by race, we observed that it was stronger for White, Black and Asian men than for Hispanic men (Asian race, aOR 0.83, 95\% CI 0.76-0.91; Black race: aOR 0.92, 95\% CI 0.88-0.96; Hispanic race: aOR 0.95, 95\% CI 0.89-1.02; White race: aOR 0.90, 95\% CI 0.87-0.92). Similar gender-based differerences in LOS were observed when a visit resulted in an admission (Asian race, aOR 0.86, 95\% CI 0.75-0.98; Black race: aOR 0.84, 95\% CI 0.79-0.90; Hispanic race: aOR 0.90, 95\% CI 0.89-1.00; White race: aOR 0.89, 95\% CI 0.88-0.93). After multiple-hypothesis correction we observed no association between LOS and insurance status, or between LOS and the use of a primary language other than English.

\begin{table}[h]
    \centering
\footnotesize
    \caption{\label{tab:wait-time-25}Adjusted odds ratios describing the effect of demographic variables on the likelihood of a very long stay (i.e. top-quartile) before admission (left) or discharge (right), relative to a specified reference group. P-values were adjusted using Bonferroni correction. Significance codes: \(p<0.05 (*); p < 0.01 (**); p < 0.001 (***)\).}
    \begin{tabular}{lc|cccc}
 & & \multicolumn{4}{c}{Very long stay before}\\
          &&  \multicolumn{2}{c}{Admission to hospital}&  \multicolumn{2}{c}{Discharge from ED}\\\hline
          Gender&Female&  1&  NA (ref)&  1& NA (ref)\\
          &Male&  0.9 (0.88-0.93)&  ***&  0.97 (0.95-0.99)& \\
          &&  &  &  & \\
          Race&White&  1&  NA (ref)&  1& NA (ref)\\
          &Asian&  1.13 (1.02-1.25)&  &  0.95 (0.88-1.02)& \\
          &Other&  1.06 (0.97-1.15)&  &  0.91 (0.85-0.96)& \\
          &Unknown&  0.62 (0.46-0.84)&  *&  1.13 (0.89-1.44)& \\
          &Black&  1.11 (1.06-1.16)&  ***&  1.01 (0.98-1.04)& \\
  &Hispanic/Latino& 1.06 (0.98-1.16)& & 0.99 (0.94-1.03)&\\
  &American Indian& 0.93 (0.64-1.37)& & 1.24 (0.89-1.74)&\\
  &Native Hawaiian& 0.83 (0.44-1.57)& & 0.83 (0.53-1.3)&\\
  && & & &\\
  Age&18-29& 1& NA (ref)& 1&NA (ref)\\
  &30-39& 1.13 (1.03-1.23)& **& 1.31 (1.26-1.37)&***\\
  &40-49& 1.22 (1.11-1.34)& ***& 1.67 (1.59-1.76)&***\\
          &50-59&  1.11 (1-1.22)&  *&  1.8 (1.71-1.89)& ***\\
 & 60-69& 1.1 (0.99-1.23)& & 1.99 (1.87-2.11)&***\\
 & 70-79& 1.12 (0.93-1.34)& & 2.15 (1.99-2.32)&***\\
 & 80-89& 1.03 (0.8-1.33)& & 2.32 (2.09-2.58)&***\\
 & 90+& 1.36 (0.91-2.04)& & 2.06 (1.75-2.44)&***\\
 & & & & &\\
 Primary language& English& 1& NA (ref)& -&-\\
 & Other& 1.06 (1.00-1.12)& & -&-\\
 & & & & &\\
 Insurance& Medicare& 1& NA (ref)& -&-\\
 & Medicaid& 1.03 (0.98-1.08)& & -&-\\
 & Private& 0.91 (0.87-0.95)& & -&-\\
 & Other/NA& 0.89 (0.80-0.99)& & -&-\\
 \hline
    \end{tabular}
\end{table}

Increasing age was strongly associated with increased LOS before discharge, and every age group over 30 was likelier to experience long or very long stays before be sent home than the under-30s if not admitted (Tables \ref{tab:wait-time-50} and \ref{tab:wait-time-25}, right columns). Age also increased the likelihood of a long stay before admission, relative to those under 30; these results were not significant for patients aged 80 or older.

Black patients were more likely than White patients to experience long stays before admission (aOR 1.14, 95\% CI 1.10-1.18), even after adjustment for chief complaint, insurance status, primary language, and eventual diagnosis (as measured by Charlson score). Black patients were still at increased risk of higher LOS when this was measured by a `very long' stay before admission, i.e. in the top quartile (see Table \ref{tab:wait-time-25}).

\begin{figure}[htb!]
\centering
\includegraphics[width=0.7\linewidth]{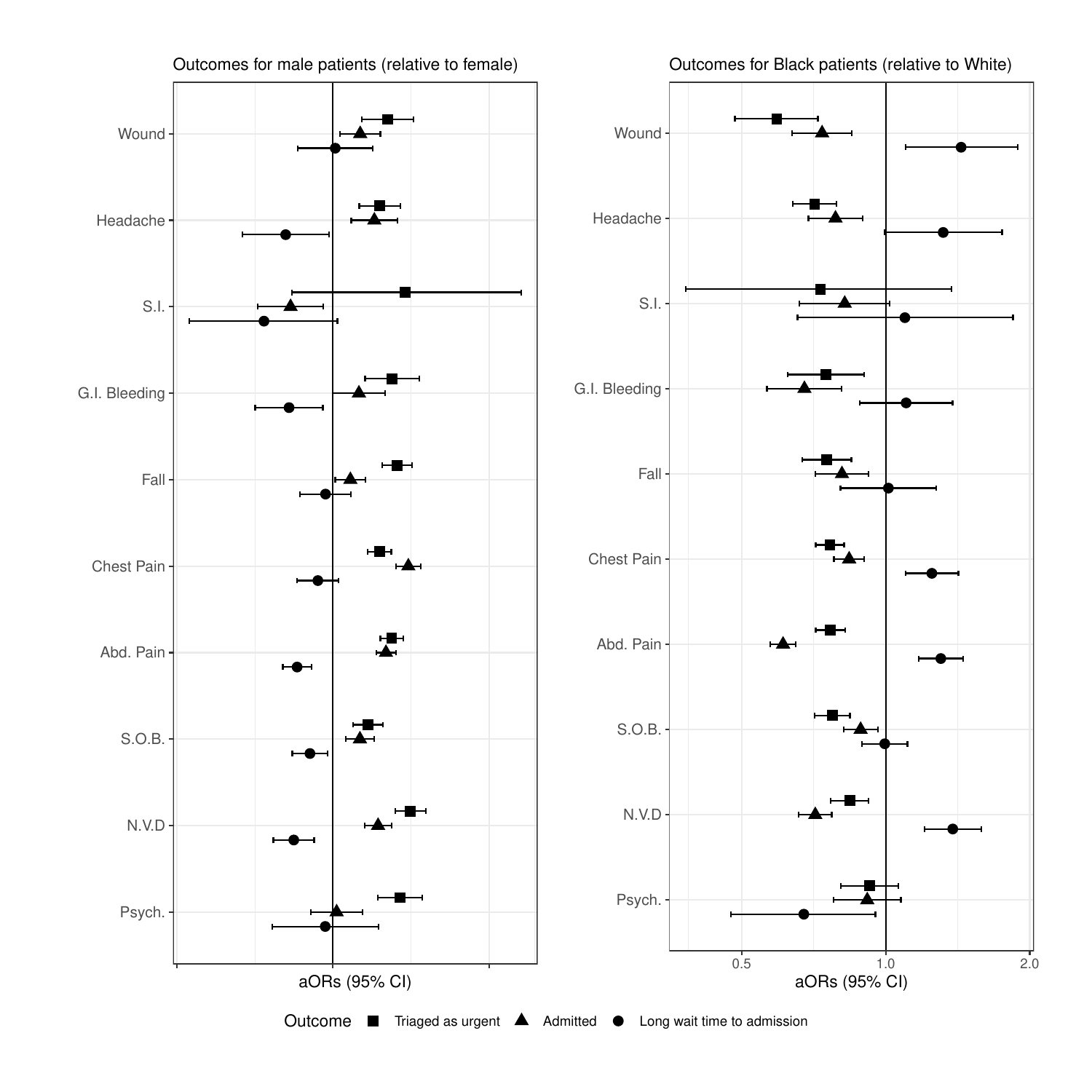}
\caption{Adjusted odds ratios for the effect of Black race on each of three outcomes, relative to White race, broken down by complaint. Odds ratios for length of stay before admission are additionally adjusted for insurance status, marital status, and primary language. S.I. = self-harm, suicidal ideation or suicide attempt, G.I. = gastrointestinal, S.O.B = shortness of breath, N.V.D = nausea, vomiting and diarrhea, `psych'= mental health conditions other than S.I., e.g. anxiety, depression, hallucinations.}
\label{fig:complaint}
\end{figure}

To further investigate this disparity, we calculated adjusted odds ratios for Black race (relative to White race) and male patients  (relative to female) presenting with each of 10 selected `chief complaints' (see Figure \ref{fig:complaint}. These were chosen to cover a wide range of reasons for an ED visit, ranging from the possibly-mild (e.g. headache) to the likely-severe (wound, suicidality, chest pain). Black patients were \textit{less} likely than White patients to experience wait times for psychiatric conditions, in line with previous findings \cite{Schrader2013}. Racial differences were particularly stark for chest and abdominal pain, which is self-reported and susceptible to racial differences in the tendency of clinical staff to believe a patient's account of their own symptoms. Alone of all conditions we investigated, men were less likely than women to be admitted for self-harm and suicide attempts. 

\section{Discussion and Conclusion}

Using a large, single-site database from the ED of a large academic medical center, we examined the relationship between the demographic characteristics reported by a patient during their ED stay and certain ED outcomes: being triaged as `urgent', admitted to hospital, or experiencing long stays before admission and discharge. Female, Black, Asian, and Hispanic patients, as well as those of `other' non-White races, are significantly less likely to be triaged as urgent, or to be admitted to hospital once acuity was accounted for.

We note that, because information on insurance was only recorded when a patient's visited resulted in admission, results for triage acuity and admission decisions do not account for differences in insurance status, nor in marital status or primary language. However, when we \textit{can} account for these differences (i.e. when comparing LOS between patients admitted to hospital), we found that female patient are still at greater risk of worse outcomes (i.e. long stays) than male patients. This may partly be due to the effect of unmeasured confounders, such as income, medical history, or the presence of a patient advocate (i.e. a spouse or child who may be better able to communicate a patient's symptoms). Differences in triage acuity may also arise from differences in the symptoms of acute medical conditions (e.g. heart attacks) between men and women, though gender-based differences persist in admission rates and wait times, where eventual diagnoses are controlled for using Charlson scores. Previous research has demonstrated that men are frequently offered more advanced medical treatment than women for the same symptoms, and that medical providers may be more likely to interpret the complaints of female patients as psychological or psychosomatic \cite{Hamberg2008}. Qualitative and mixed-methods studies are necessary to clarify the extent to which gender-based outcome differences arise from provider biases, or from differences in the communication strategies employed by male and female patients. Such studies should also record the gender of the triage nurse and whether or not it differs from the patient's gender.

\begin{figure}[htb!]
\centering
\includegraphics[width=0.7\linewidth]{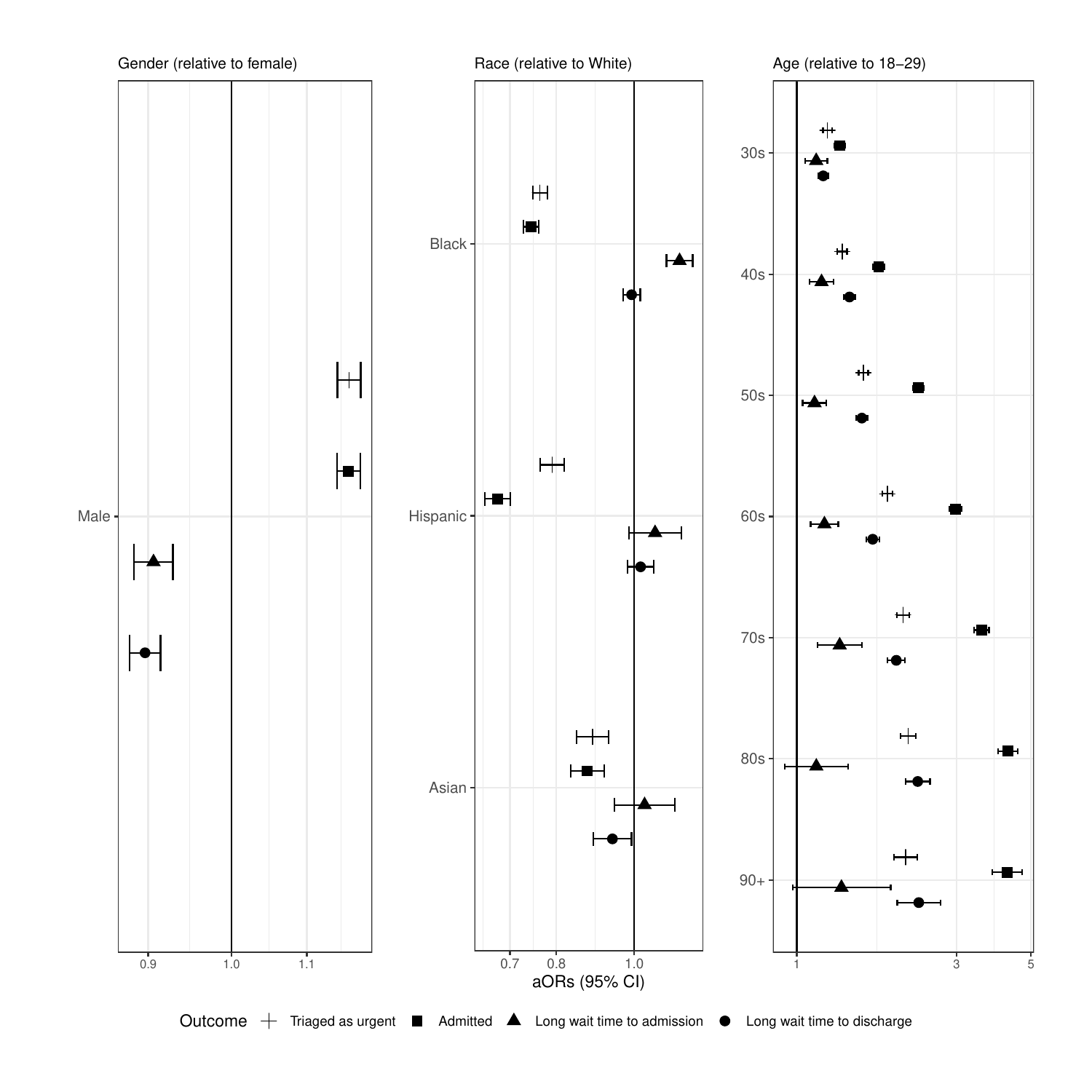}
\caption{Adjusted odds ratios for the effect of each exposure on each of the four outcomes under study. Odds ratios for length of stay before admission are additionally adjusted for insurance status, marital status, and primary language.}
\label{fig:ors}
\end{figure}

We also find that Black patients experience worse outcomes relative to White patients (see Figure \ref{fig:ors}), in line with previous literature \cite{Park2009}, even when presenting with the same conditions and when their stays result in similar diagnoses. These results may also be partly attributable to provider bias, which may in turn create obstacles to communication, as evidenced by mixed-methods work demonstrating that providers are likelier to interrupt patients whose race differs from their own \cite{Aysola2022}. Unlike female gender, Black race (relative to White race) increased a patient's LOS when waiting to be admitted, but not when waiting to be discharged. 

We note certain limitations in our quantification of patient race. Our analysis used only coarse racial categories (i.e. `Asian' rather than `Korean' or `Chinese'), as this information was only available for a minority of patients. Coarse-graining may obscure outcome differences within racial groups \cite{Movva2023}, and further granularity may clarify the source of these disparities. Our exclusion procedures may also bias our study if data is more likely to be missing for patients of marginalised races or ethnicities, or for patients with specific conditions, who may bypass triage entirely. Higher sample sizes are also required to achieve balance between Native and White populations and clarify the differences in outcomes between these groups. Due to the use of a single collapsed `race' field, we were also unable to clarify whether those who reported White or Black race were in fact non-Hispanic.

We found that age up until 80 generally results in a higher likelihood of being triaged as urgent or admitted to hospital, but (accounting for insurance and marital status) a longer stay in the ED, relative to those under 30. This may reflect differences between patient presentations not captured by the adjustment factors we used in this study.  For example, patients under 30 may arrive with a single acute complaint, whereas older patients may have a number of interacting conditions, complicating the decisions as to whether or not to admit them (leading to longer stays before discharge) and, if so, into which department (leading to longer stays before admission). We note that we were not able to distinguish between the time to an admission decision and the time a patient spent waiting for a bed once a decision had been made. Relatedly, we were also unable to account for the degree of overcrowding within the ED (outside of using time of day as a proxy).

During this study, we made every effort to adjust for factors associated with a patient's condition, such as vital signs at triage, whether or not a patient was able to report a pain score, and whether a patient arrived by helicopter or ambulance. However, we have not eliminated all possible sources of confounding. 
Our results also do not necessarily extend beyond the hospital under study, to pediatric patients, or to years beyond 2019 (i.e. to the experience of patients during the COVID-19 pandemic). Multi-site studies using more recent data are necessary to establish the generalisability of these findings. However, these findings highlight important sources of possible bias in widely available public datasets, and emphasise that researchers should be alert to these biases when designing tools for clinical intervention. Standardised care pathways which reduce provider discretion, as well as bias awareness training programs, may also be necessary to increase the fairness of outcomes between patient groups.

\section{Acknowledgments}
Research reported in this publication was supported by the National Library Of Medicine of the National Institutes of Health under Award Number R01LM014300. 
The content is solely the responsibility of the authors and does not necessarily represent the official views of the National Institutes of Health.

\bibliography{sample}

\end{document}